\documentclass[twocolumn,showpacs,preprintnumbers,amsmath,amssymb, showkeys]{revtex4}

\usepackage{graphicx}
\usepackage{dcolumn}
\usepackage{bm}
\usepackage{epsf}

\begin{document}


\title{Correlation induced switching of local spatial charge
distribution in two-level system}

\author{P.\,I.\,Arseyev}
 \altaffiliation{ars@lpi.ru}
\author{N.\,S.\,Maslova}%
 \email{spm@spmlab.phys.msu.ru}
\author{V.\,N.\,Mantsevich}
 \altaffiliation{vmantsev@spmlab.phys.msu.ru}
\affiliation{%
 P.N. Lebedev Physical institute of RAS, 119991, Moscow, Russia\\~\\
 Moscow State University, Department of  Physics,
119991 Moscow, Russia
}%

\date{\today }
6 pages, 4 figures
\begin{abstract}
We present theoretical investigation of spatial charge distribution
in the two-level system with strong Coulomb correlations by means of
Heisenberg equations analysis for localized states total electron
filling numbers taking into account pair correlations of local
electron density. It was found that tunneling current through
nanometer scale structure with strongly coupled localized states
causes Coulomb correlations induced spatial redistribution of
localized charges. Conditions for inverse occupation of two-level
system in particular range of applied bias caused by Coulomb
correlations have been revealed. We also discuss possibility of
charge manipulation in the proposed system.
\end{abstract}

\maketitle

\section{Introduction}

Investigation of tunneling properties of interacting impurity
complexes in the presence of Coulomb correlations is one of the most
important problems in the physics of nanostructures. Tunneling
current changes localized states electron filling numbers as a
result-the spectrum and electron density of states are also modified
due to Coulomb interaction of localized electrons. Moreover the
charge distribution in the vicinity of such complexes can be tuned
by changing the parameters of the tunneling contact. Self-consistent
approach based on Keldysh diagram technique have been successfully
used to analyze non-equilibrium effects and tunneling current
spectra in the system of two weakly coupled impurities (when
coupling between impurities is smaller than tunneling rates between
energy levels and tunneling contact leads) in the presence of
Coulomb interaction \cite{Keldysh}. In the mean-field approximation
for mixed valence regime the dependence of electron filling numbers
on applied bias voltage and the behaviour of tunneling current
spectra have been analyzed in \cite{Arseyev}.

Electron transport even through a single impurity in the Coulomb
blockade and the Kondo regime \cite{Kondo} have been studied
experimentally and is up till now under theoretical investigation
\cite{Goldin}-\cite{Kikoin}.  As tunneling coupling is not
negligable the impurity charge is not the discrete value and one has
to deal with impurity electron filling numbers (which now are
continuous variables) determined from kinetic equations.

Analyzing non-equilibrium tunneling processes through coupled
impurities one can reveal switching on and off of magnetic regime
(electron filling numbers in the localized states for opposite spins
are equal) on each impurity atom at particular range of applied bias
voltage \cite{Arseyev}.

In the present work we consider the opposite case when coupling
between localized electron states strongly exceeds tunneling
transfer rates. This situation can be experimentally realized when
several impurity atoms or surface defects are situated at the
neighboring lattice sites, so coupling between their electronic
states can strongly exceeds the interaction of this localized states
with continuous spectrum Fig.\ref{Fig.1}. Another possible
realization is two interacting quantum dots on the sample surface
weakly connected with the bulk states. Such systems can be described
by the model including several electron levels with Coulomb
interaction between localized electrons. If the distance between
impurities is smaller than localization radius, strong enough
correlation effects arise which modify the spectrum of the whole
complex. Electronic structure of such complexes can be tuned as by
external electric field which changes the values of single particle
levels as by electron correlations of localized electronic states.
One can expect that tunneling current induces non-equilibrium
spatial redistribution of localized charges and gives possibility of
local charge density manipulation strongly influenced by Coulomb
correlations. In some sense these effects are similar to the
$"$co-tunneling$"$ observed in \cite{Feigel'man},
\cite{Beloborodov}. Moreover Coulomb interaction of localized
electrons can be responsible for inverse occupation of localized
electron states. These effects can be clearly seen when single
electron levels have different spatial symmetry.

To understand such correlation induced $"$charge$"$ switching it's
sufficient to analyze Heisenberg equations for localized states
total electron filling numbers taking into account pair correlations
of local electron density \cite{Maslova}. If one is interested in
kinetic properties and changes of local charge density for the
applied bias range higher than the value of energy levels tunneling
broadening modification of initial density of states due to the
Kondo effect can be neglected. In this case for the finite number of
localized electron levels one can obtain closed system of equations
for electron filling numbers and their higher order correlations.

\section{The suggested model}

\begin{figure} [h!]
\includegraphics[width=70mm]{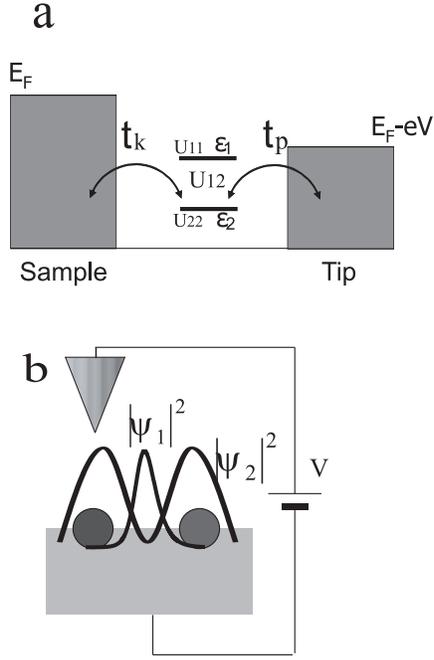}
\caption{ a). Energy diagram of two-level system and b). Schematic
spatial diagram of experimental realization. Coulomb energy $U_{12}$
correspond to the interaction between electrons on different energy
levels.} \label{Fig.1}
\end{figure}

We shall analyze tunneling through the two-level system with Coulomb
interaction Fig.\ref{Fig.1}. The model system can be described by
the Hamiltonian $\hat{H}$.

\begin{eqnarray}
\Hat{H}&=&\sum_{i\sigma}\varepsilon_{i}n_{i\sigma}+\sum_{k\sigma}\varepsilon_{k}c_{k\sigma}^{+}c_{k\sigma}+\sum_{p\sigma}\varepsilon_{p}c_{p\sigma}^{+}c_{p\sigma}+\nonumber\\
&+&\sum_{ij\sigma\sigma^{'}}U_{ij}^{\sigma\sigma^{'}}n_{i\sigma}n_{j\sigma^{'}}+\sum_{ki\sigma}t_{k}c_{k\sigma}^{+}c_{i\sigma}+\nonumber\\
&+&\sum_{pi\sigma}t_{p}c_{p\sigma}^{+}c_{i\sigma}+h.c.\
\end{eqnarray}

Indices $k$ and $p$ label continuous spectrum states in the left
(sample) and right (tip) leads of tunneling contact respectively.
$t_{k(p)}$- tunneling transfer amplitudes between continuous
spectrum states and two-level system with elctron levels
$\varepsilon_i$.
 Operators
$c_{k(p)}^{+}/c_{k(p)}$ correspond to electrons
creation/annihilation in the continuous spectrum states $k(p)$.
$n_{i\sigma}=c_{i\sigma}^{+}c_{i\sigma}$-two-level system electron
filling numbers, where operator $c_{i\sigma}$ destroys electron with
spin $\sigma$ on the energy level $\varepsilon_i$.
$U_{ij}^{\sigma\sigma^{'}}$ is the on-site Coulomb repulsion of
localized electrons.

Tunneling current through the two-level system can be written in the
terms of electron creation/annihilation operators as:

\begin{eqnarray}
I=I_{k\sigma}=\sum_{k\sigma}\dot{n}_{k\sigma}=\sum_{ki\sigma}t_{k}(<c_{k\sigma}^{+}c_{i\sigma}>-<c_{i\sigma}^{+}c_{k\sigma}>)
\end{eqnarray}

Let us consider $\hbar=1$ elsewhere, so motion equation for the
electron operators product $c_{k\sigma}^{+}c_{i\sigma}$ can be
written as:

\begin{eqnarray}
i\frac{\partial c_{k\sigma}^{+}c_{i\sigma}}{\partial
t}&=&(\varepsilon_i-\varepsilon_k)\cdot
c_{k\sigma}^{+}c_{i\sigma}+U_{ii}n_{i-\sigma}\cdot
c_{k\sigma}^{+}c_{i\sigma}+\nonumber\\
&+&U_{ij}(n_{j\sigma}+n_{j-\sigma})\cdot
c_{k\sigma}^{+}c_{i\sigma}-t_{k}\cdot (n_{i\sigma}-\widehat{f}_{k})+\nonumber\\
&+&\sum_{k^{'}\neq
k}t_{k^{'}}c_{k\sigma}^{+}c_{k^{'}\sigma}+\sum_{i\neq
j}t_{k}c_{j\sigma}^{+}c_{i\sigma}=0 \label{1}
\end{eqnarray}

where

\begin{eqnarray}
\widehat{f}_{k}=c_{k\sigma}^{+}c_{k\sigma}
\end{eqnarray}

Now let us also consider that $n_{i\sigma}^{2}=n_{i\sigma}$.

Neglecting changes of electron spectrum and local density of states
in the tunneling contact leads due to the tunneling current flowing
we shall uncouple conduction and two-level system electron filling
numbers. After summation over $k$ one can get an equation which
describe tunneling current in the presented two-level system:

\begin{eqnarray}
I_{k\sigma}=I_{k1\sigma}+I_{k2\sigma}
\end{eqnarray}

Where expression for tunneling current $I_{k2\sigma}$ can be
obtained by changing indexes $1\leftrightarrow2$ in equation for
tunneling current $I_{k1\sigma}$ which has the form:

\begin{eqnarray}
I_{k1\sigma}&=&\Gamma_{k}\cdot\{\langle n_{1\sigma}\rangle+\sum_{j\neq i}\langle c_{j\sigma}^{+}c_{i\sigma}\rangle-\nonumber\\
&-&\langle(1-n_{1-\sigma})(1-n_{2-\sigma})(1-n_{2\sigma})\rangle\cdot f_{k}(\varepsilon_1)-\nonumber\\
&-&\langle n_{1-\sigma}(1-n_{2-\sigma})(1-n_{2\sigma})\rangle\cdot f_{k}(\varepsilon_1+U_{11})-\nonumber\\
&-&\sum_{\sigma^{'}}\langle n_{2\sigma^{'}}(1-n_{2-\sigma^{'}})(1-n_{1-\sigma})\rangle\cdot f_{k}(\varepsilon_1+U_{12})-\nonumber\\
&-&\sum_{\sigma^{'}}\langle n_{1-\sigma}n_{2\sigma^{'}}(1-n_{2-\sigma^{'}})\rangle\cdot f_{k}(\varepsilon_1+U_{11}+U_{12})-\nonumber\\
&-&\langle n_{2\sigma}n_{2-\sigma}(1-n_{1-\sigma})\rangle\cdot f_{k}(\varepsilon_1+2U_{12})-\nonumber\\
&-&\langle n_{1-\sigma}n_{2-\sigma}n_{2\sigma}\rangle\cdot
f_{k}(\varepsilon_1+U_{11}+2U_{12})\}+\nonumber\\
&+&\sum_{k^{'}\neq
k}\langle t_{k}t_{k^{'}}c_{k\sigma}^{+}c_{k^{'}\sigma}\rangle\cdot\nonumber\\
&\cdot&\{\langle\frac{(1-n_{1-\sigma})(1-n_{2-\sigma})(1-n_{2\sigma})}{\varepsilon_1-\varepsilon_k}\rangle+\nonumber\\
&+&\langle\frac{n_{1-\sigma}(1-n_{2-\sigma})(1-n_{2\sigma})}{\varepsilon_1+U_{11}-\varepsilon_k}\rangle+\nonumber\\
&+&\langle\frac{\sum_{\sigma^{'}}n_{2\sigma^{'}}(1-n_{1-\sigma})(1-n_{2-\sigma^{'}})}{\varepsilon_1+U_{12}-\varepsilon_k}\rangle+\nonumber\\
&+&\langle\frac{\sum_{\sigma^{'}}n_{1-\sigma}n_{2\sigma^{'}}(1-n_{2-\sigma^{'}})}{\varepsilon_1+U_{11}+U_{12}-\varepsilon_k}\rangle+\nonumber\\
&+&\langle\frac{n_{2-\sigma}n_{2\sigma}(1-n_{1-\sigma})}{\varepsilon_1+2U_{12}-\varepsilon_k}\rangle+\langle\frac{n_{1-\sigma}n_{2-\sigma}n_{2\sigma}}{\varepsilon_1+U_{11}+2U_{12}-\varepsilon_k}\rangle\}\nonumber\\
\label{current}
\end{eqnarray}

We shall further neglect terms $t_{k}c_{i\sigma}^{+}c_{j\sigma}$ and
$t_{k}c_{k\sigma}^{+}c_{k^{'}\sigma}$ in expression \ref{current} as
they correspond to the next order perturbation theory by the
parameter $\frac{\Gamma_{i}}{\Delta\varepsilon_{i}}$. Relaxation
rates $\Gamma_{k(p)}=\pi\cdot t_{k(p)}^{2}\cdot\nu_{0}$ are
determined by electron tunneling transitions from two-level system
to the leads $k$ (sample) and $p$ (tip) continuum states.
$\nu_{0}$-continuous spectrum density of states. Equations for
filling numbers $n_{1\sigma}$ è $n_{2\sigma}$ can be found from the
conditions:

\begin{eqnarray}
\frac{\partial n_{1\sigma}}{\partial t}=I_{k1\sigma}+I_{p1\sigma}=0\nonumber\\
\frac{\partial n_{2\sigma}}{\partial t}=I_{k2\sigma}+I_{p2\sigma}=0\
\label{filling numbers}
\end{eqnarray}

where tunneling current $I_{p\sigma}$ can be easily determined from
$I_{k\sigma}$ by changing indexes $k\leftrightarrow p$

We shall analyze the situation when Coulomb energy values are large
and condition $U_{ij}>>\varepsilon_{i/j}$ can be taken into account.
It means that if one have to calculate tunneling current through
such system it is necessary to find all pair filling numbers
correlators in the energy range $\varepsilon_i+U_{ij}$. So we retain
the terms containing $f_{k(p)}(\varepsilon_i+U_{ij})$ and neglect
all high orders correlators and pair correlators which contain
$f_{k(p)}(\varepsilon_i+U_{ij}+U_{kl})$. We consider the
paramagnetic situation $n_{i\sigma}=n_{i-\sigma}$.

Pair filling numbers correlators can be found in the following way:

\begin{eqnarray}
\langle\frac{\partial n_{i\sigma}n_{j\sigma^{'}}}{\partial
t}\rangle=\langle\frac{\partial n_{i\sigma}}{\partial
t}n_{j\sigma^{'}}\rangle+\langle\frac{\partial
n_{j\sigma^{'}}}{\partial t}n_{i\sigma}\rangle \label{correlators}
\end{eqnarray}

Let us introduce tunneling filling numbers $n^{T}(\varepsilon_i)$,
$n^{T}(\varepsilon_i+U_{ij})$ and $\tilde{n}_{ij}^{T}$ which have
the form:

\begin{eqnarray}
n^{T}(\varepsilon_i)&=&\frac{\Gamma_{k}f_{k}(\varepsilon_i)+\Gamma_{p}f_{p}(\varepsilon_i)}{\Gamma_{k}+\Gamma_{p}}\nonumber\\
n^{T}(\varepsilon_i+U_{ij})&=&\frac{\Gamma_{k}f_{k}(\varepsilon_i+U_{ij})+\Gamma_{p}f_{p}(\varepsilon_i+U_{ij})}{\Gamma_{k}+\Gamma_{p}}\nonumber\\
\tilde{n}_{ij}^{T}&=&\frac{\Gamma_{k}\tilde{f}_{kij}+\Gamma_{p}\tilde{f}_{pij}}{\Gamma_{k}+\Gamma_{p}}
\end{eqnarray}

where

\begin{eqnarray}
\tilde{f}_{kij}=f_{k}(\varepsilon_i)-f_{k}(\varepsilon_{i}+U_{ij})\
\end{eqnarray}

As we consider that $n_{i\sigma}=n_{i-\sigma}$, let us also consider
$\langle n_{i\sigma}n_{j\sigma}\rangle=\langle
n_{i\sigma}n_{j-\sigma}\rangle$. So a system of equations for pair
correlators $K_{11}\equiv \langle n_{1\sigma}n_{1-\sigma}\rangle$,
$K_{22}\equiv \langle n_{2\sigma}n_{2-\sigma}\rangle$ and
$K_{12}\equiv \langle n_{1\sigma}n_{2\sigma}\rangle$ for large
Coulomb energies $U_{ij}>>\varepsilon_{i/j}$ has the form:

\begin{eqnarray}
\begin{pmatrix}a_{11} &&  a_{12} && a_{13}\\
a_{21} &&  a_{22} && a_{23}\\
a_{31} &&  a_{32} && a_{33}\end{pmatrix}\times
\begin{pmatrix}K_{11}\\K_{22}\\K_{12}\end{pmatrix}=F
\label{matrix}
\end{eqnarray}

where

\begin{eqnarray}
a_{11}&=&a_{23}=1\nonumber\\
a_{13}&=&a_{21}=0\nonumber\\
a_{12}&=&2n^{T}(\varepsilon_1+U_{11})\nonumber\\
a_{22}&=&2n^{T}(\varepsilon_2+U_{22})\
\end{eqnarray}

\begin{eqnarray}
a_{31}&=&\frac{1}{2}n^{T}(\varepsilon_2+U_{12})\nonumber\\
a_{32}&=&1+\frac{1}{2}n^{T}(\varepsilon_1+U_{12})+\frac{1}{2}n^{T}(\varepsilon_2+U_{12})\nonumber\\
a_{33}&=&\frac{1}{2}n^{T}(\varepsilon_1+U_{12})\
\end{eqnarray}

and

\begin{eqnarray}
F=\begin{pmatrix} n^{T}(\varepsilon_1+U_{11})\cdot n_{1\sigma}\\ n^{T}(\varepsilon_2+U_{22})\cdot n_{2\sigma}\\ \frac{1}{2}n^{T}(\varepsilon_1+U_{12})\cdot n_{2\sigma}+\frac{1}{2}n^{T}(\varepsilon_2+U_{12})\cdot n_{1\sigma}\\
\end{pmatrix}
\end{eqnarray}

Equations which determine two-level system filling numbers
immediately follows from the system \ref{matrix}:

\begin{figure*} [t]
\includegraphics[width=160mm]{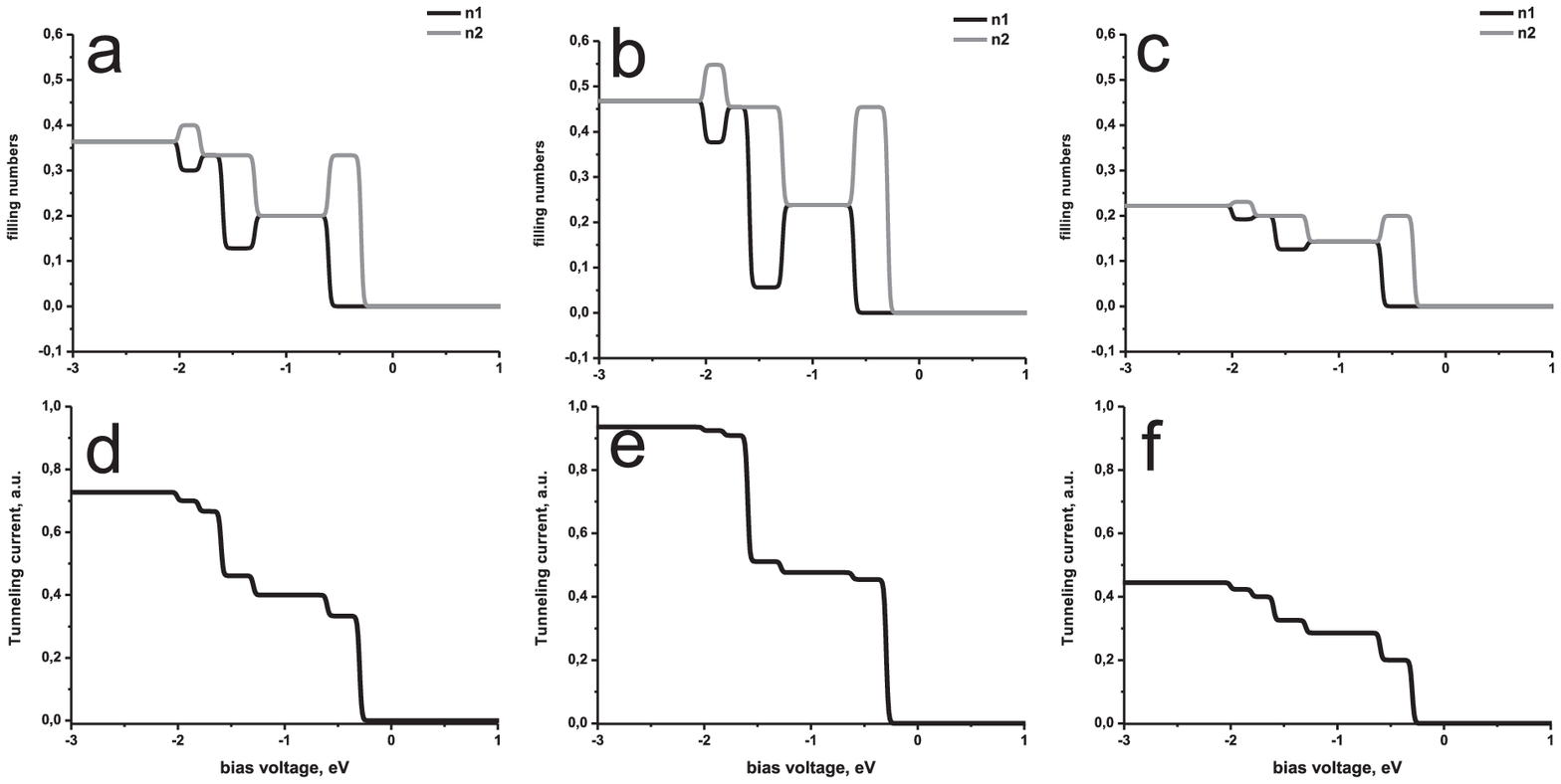}
\caption{Two-level system filling numbers a).-c). and tunneling
current d).-f). as a function of applied bias voltage in the case
when both energy levels are situated above the sample Fermi level.
Parameters $\epsilon_{1}=0.6$, $\epsilon_{2}=0.3$, $U_{12}=1.0$,
$U_{11}=1.4$, $U_{22}=1.5$ are the same for all the figures.
a),d).$\Gamma_{k}=0.01$, $\Gamma_{p}=0.01$; b),e).$\Gamma_{k}=0.05$,
$\Gamma_{p}=0.01$; c),f).$\Gamma_{k}=0.01$, $\Gamma_{p}=0.03$.}
\label{Fig.2}
\end{figure*}

\begin{eqnarray}
n_{1\sigma}&\cdot&
(1+\tilde{n}_{11}^{T})+n_{2\sigma}\cdot2\tilde{n}_{12}^{T}-K_{22}\cdot
(n^{T}(\varepsilon_1)-2n^{T}(\varepsilon_1+U_{12}))+\nonumber\\
&+&2\cdot
K_{12}\cdot(-n^{T}(\varepsilon_1)+n^{T}(\varepsilon_1+U_{11})+n_{T}(\varepsilon_1+U_{12}))=\nonumber\\
&=&n^{T}(\varepsilon_1)\nonumber\\
n_{2\sigma}&\cdot&
(1+\tilde{n}_{22}^{T})+n_{1\sigma}\cdot2\tilde{n}_{21}^{T}-K_{11}\cdot
(n^{T}(\varepsilon_2)-2n^{T}(\varepsilon_2+U_{12}))+\nonumber\\
&+&2\cdot
K_{12}\cdot(-n^{T}(\varepsilon_2)+n^{T}(\varepsilon_2+U_{22})+n_{T}(\varepsilon_2+U_{12}))=\nonumber\\
&=&n^{T}(\varepsilon_2) \label{2}
\end{eqnarray}

And finally expression for tunneling current has the form:

\begin{eqnarray}
I_{k1\sigma}&=&\Gamma_{k}\cdot\{\langle n_{1\sigma}\rangle-
(1-\langle n_{1\sigma}\rangle-2\langle n_{2\sigma}\rangle+K_{22}+2K_{12})\cdot\nonumber\\
&\cdot&f_{k}(\varepsilon_1)-
(\langle n_{1\sigma}\rangle-2K_{12})\cdot f_{k}(\varepsilon_1+U_{11})-\nonumber\\
&-&2\cdot(\langle n_{2\sigma}\rangle-K_{12}-K_{22})\cdot
f_{k}(\varepsilon_1+U_{12})\ \label{current_final}
\end{eqnarray}

Let us also mention two extreme cases. The first one when all
Coulomb energies are extremely large $U_{ij}\rightarrow\infty$. In
this situation expressions for filling numbers will have the
following form:

\begin{eqnarray}
n_{1\sigma}=\frac{n^{T}(\varepsilon_1)\cdot(1-n^{T}(\varepsilon_2))}{(1+n^{T}(\varepsilon_1))\cdot(1+n^{T}(\varepsilon_2))-4\cdot n^{T}(\varepsilon_1)\cdot n^{T}(\varepsilon_2)}\nonumber\\
n_{2\sigma}=\frac{n^{T}(\varepsilon_2)\cdot(1-n^{T}(\varepsilon_1))}{(1+n^{T}(\varepsilon_1))\cdot(1+n^{T}(\varepsilon_2))-4\cdot n^{T}(\varepsilon_1)\cdot n^{T}(\varepsilon_2)}\nonumber\\
\end{eqnarray}

And the second one is when energy levels are generated, for example
due to the orbital quantum number
$\varepsilon_1=\varepsilon_2=\varepsilon$ and consequently
$U_{ij}=U$. In this case filling numbers have the form:

\begin{eqnarray}
n_{\sigma}=\frac{n^{T}(\varepsilon)}{1+3\cdot n^{T}(\varepsilon)}
\end{eqnarray}

\begin{figure*} [t]
\includegraphics[width=160mm]{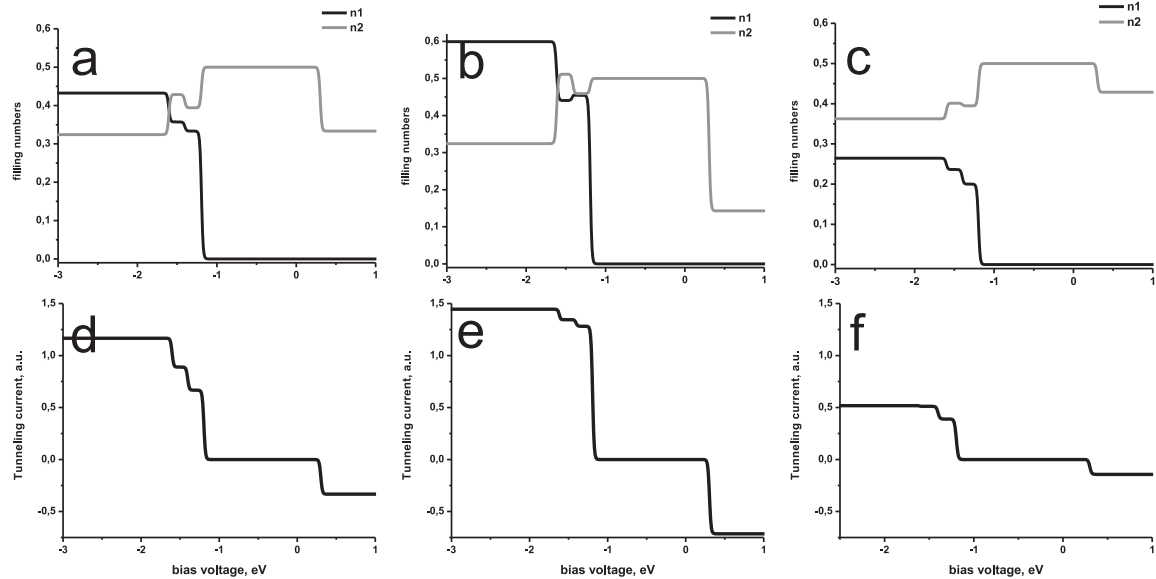}
\caption{Two-level system filling numbers a).-c). and tunneling
current d).-f). as a function of applied bias voltage in the case
when one energy level is situated above and another one below the
sample Fermi level. Parameters $\epsilon_{1}=0.2$,
$\epsilon_{2}=-0.3$, $U_{12}=1.0$, $U_{11}=1.4$, $U_{22}=1.7$ are
the same for all the figures. a),d).$\Gamma_{k}=0.01$,
$\Gamma_{p}=0.01$; b),e).$\Gamma_{k}=0.05$, $\Gamma_{p}=0.01$;
c),f).$\Gamma_{k}=0.01$, $\Gamma_{p}=0.03$.} \label{Fig.3}
\end{figure*}

\section{Main results and discussion}
\begin{figure*} [t]
\includegraphics[width=160mm]{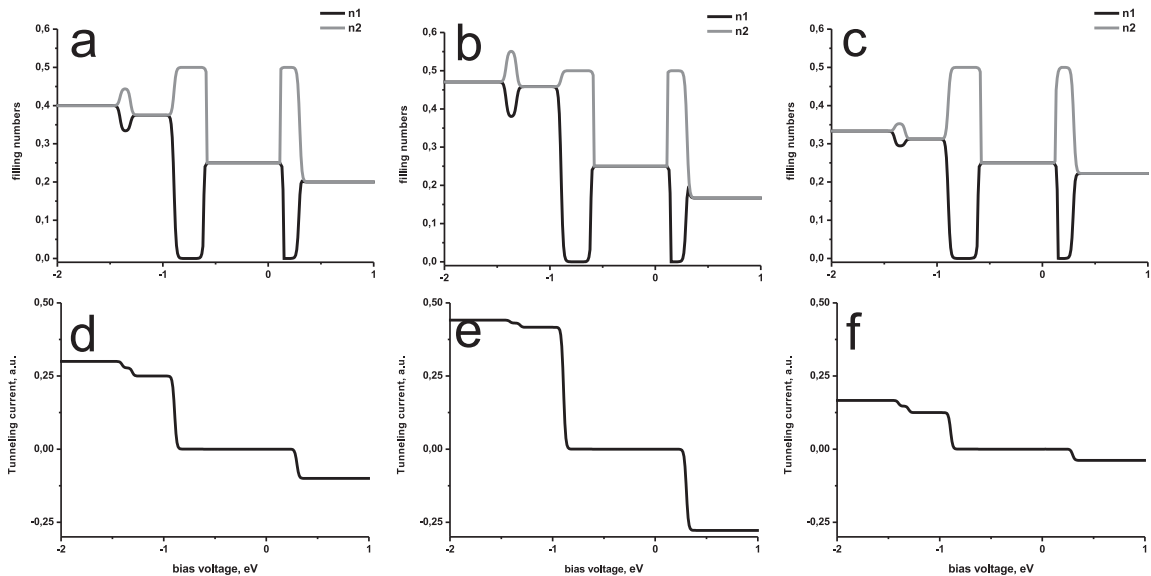}
\caption{Two-level system filling numbers a).-c). and tunneling
current d).-f). as a function of applied bias voltage in the case
when both energy levels are situated below the sample Fermi level.
Parameters $\epsilon_{1}=-0.1$, $\epsilon_{2}=-0.3$, $U_{12}=1.0$,
$U_{11}=1.5$, $U_{22}=1.6$ are the same for all the figures.
a),d).$\Gamma_{k}=0.01$, $\Gamma_{p}=0.01$; b),e).$\Gamma_{k}=0.05$,
$\Gamma_{p}=0.01$; c),f).$\Gamma_{k}=0.01$, $\Gamma_{p}=0.03$.}
\label{Fig.4}\end{figure*}

The behaviour of non-equilibrium electron filling numbers with
changing of applied bias and tunneling conductivity characteristics
obtained from equations (\ref{current}) and
(\ref{matrix})-(\ref{current_final}) are depicted on
Fig.\ref{Fig.2}-Fig.\ref{Fig.4}.

We consider different experimental realizations: both energy levels
are situated above the  sample Fermi level (Fig.\ref{Fig.2}); both
levels below sample Fermi level (Fig.\ref{Fig.4}) and one of the
energy levels is located above the Fermi level and another one below
the Fermi level (Fig.\ref{Fig.3}). From the obtained results one can
clearly see charge redistribution between two electron states with
changing of applied bias voltage (Fig.\ref{Fig.2}-\ref{Fig.4}).

When both levels are situated above (Fig.\ref{Fig.2}) or below
(Fig.\ref{Fig.4}) the sample Fermi level one can clearly reveal two
possibilities for charge distribution in the two-level system. The
first one corresponds to the case when local charge is mostly
accumulated on the lower electron level $n_{1}<n_{2}$
($\varepsilon_2<eV<\varepsilon_1$,
$\varepsilon_2+U_{12}<eV<\varepsilon_1+U_{12}$ and
$\varepsilon_2+U_{22}<eV<\varepsilon_1+U_{11}$ on Fig.\ref{Fig.2}
and Fig.\ref{Fig.4}). The second one deals with the case when charge
is localized on both levels equally $n_{1}=n_{2}$
($\varepsilon_1<eV<\varepsilon_2+U_{12}$,
$\varepsilon_1+U_{12}<eV<\varepsilon_2+U_{22}$ and
$\varepsilon_1+U_{11}<eV$ on Fig.\ref{Fig.2} and Fig.\ref{Fig.4}).

Coulomb correlation induced sudden jumps down and up of each level
electron filling numbers at certain values of applied bias are
clearly seen.

So if electron states have essentially different symmetry one can
expect charge accumulation in various spatial areas and thus the
possibility of local charge manipulation appears.

When both electron energies are situated below the sample Fermi
level upper electron state become empty ($n_1=0$) for two ranges of
applied bias voltage ($\varepsilon_2<eV<\varepsilon_1$ and
$\varepsilon_2+U_{12}<eV<\varepsilon_1+U_{12}$ )(Fig.\ref{Fig.4}).

Described peculiarities take place for all the ratios between
tunneling transfer rates $\Gamma_k$ and $\Gamma_p$.

The other interesting effect is the possibility of inverse
occupation of the two-level system due to Coulomb interaction in
special range of applied bias (Fig.\ref{Fig.3}). In the absence of
Coulomb interaction difference of electron filling numbers is
determined by electron tunneling rates
$n_{1}-n_{2}\sim\gamma_{k1}\gamma_{p2}-\gamma_{p1}\gamma_{k2}$. So
without Coulomb interaction, for $\gamma_{k(p)1}=\gamma_{k(p)2}$,
difference of the two levels occupation numbers turns to zero.
Coulomb interaction of localized electrons in the two-level system
results in inverse occupation of two levels at the high range of
applied bias voltage. This situation is clearly demonstrated on the
Fig.\ref{Fig.3}.

It is clearly evident (Fig.\ref{Fig.3}) that when applied bias
doesn't exceed value $\varepsilon_1+U_{12}$ all the charge is
localized on the lower energy level ($n_{1}=0$). With the increasing
of applied bias inverse occupation takes place and charge localized
in the system redistributes. Local charge is mostly accumulated on
the upper level when applied bias value exceed
$\varepsilon_1+U_{11}$. Two-level system demonstrates such behaviour
if the tunneling contact is symmetrical (Fig.\ref{Fig.3}a) or when
system strongly coupled with tunneling contact lead k (sample)
(Fig.\ref{Fig.3}b). We have not found inverse occupation when
two-level system mostly coupled with tunneling contact lead p (tip)
(Fig.\ref{Fig.3}c). In this case with the increasing of applied bias
upper electron state charge also increases but local charge continue
being mostly accumulated on the lower electron state.

We also analyzed tunneling current as a function of applied bias
voltage for different level's positions
(Fig.\ref{Fig.2}-Fig.\ref{Fig.4}d-f) . Tunneling current amplitudes
presented in this work are normalized on $2\Gamma_k$ elsewhere. For
all the values of the system parameters tunneling current dependence
on applied bias has step structure. Height and length of the steps
depend on the parameters of the tunneling contact (tunneling
transfer rates and values of Coulomb energies). When both energy
levels are above the Fermi level one can find six steps in tunneling
current (Fig.\ref{Fig.2}d-f). If both levels are situated below the
Fermi level there are four steps in tunneling current
(Fig.\ref{Fig.4}d-f) and the upper electron level doesn't appear as
a step in current-voltage characteristics but charge redistribution
takes place due to Coulomb correlations. One can also reveal four
steps in the case when only lower energy level is situated below the
Fermi level (Fig.\ref{Fig.3}d-f).

\section{Conclusion}
Tunneling through the two-level system with strong coupling between
localized electron states was analyzed by means of Heisenberg
equations for localized states total electron filling numbers taking
into account high order correlations of local electron density.
Various electron levels location relative to the sample Fermi level
in symmetric and asymmetric tunneling contact were investigated.

We revealed that charge redistribution between electron states takes
place in suggested model when both electron levels are situated
above or below the sample Fermi level. Charge redistribution is
governed by Coulomb correlations. Moreover with variation of Coulomb
interaction of localized electrons one can find the bias range of
the two-level system inverse occupation when electron levels are
localized on the opposite sites of the sample Fermi level.

This work was partially supported by RFBR grants.


\pagebreak

\end{document}